\documentclass[twocolumn,showpacs,preprintnumbers,amsmath,amssymb,superscriptaddress,aps,prd]{revtex4-1}

\usepackage{graphicx}
\usepackage{dcolumn}
\usepackage{bm}
\usepackage{amsmath,amsfonts,amssymb}
\usepackage{graphicx,epic,eepic}
\usepackage{subfigure}
\usepackage{graphics}
\usepackage[dvips]{color}
\usepackage{epsfig}
\usepackage{units}
\usepackage{mathrsfs}

\def \beq {\begin{equation}}
\def \eeq {\end{equation}}

\begin{document}

\title{Challenging the weak cosmic censorship conjecture with charged quantum particles}

\author{Maur\'icio Richartz}
\email{richartz@ift.unesp.br}
\affiliation{ Instituto de F\'isica Te\'orica, Universidade Estadual Paulista,
Rua Dr. Bento Teobaldo Ferraz, 271 - Bl. II, 01140-070, S\~ao Paulo, SP, Brazil.}
\author{Alberto Saa}
\email{asaa@ime.unicamp.br}
\affiliation{
Departamento de Matem\'atica Aplicada,
 UNICAMP,  13083-859 Campinas, SP, Brazil.}

\begin{abstract}
Motivated by the recent attempts to violate the weak cosmic censorship conjecture
for near-extreme black-holes,
 we consider  the possibility of overcharging a near-extreme Reissner-Nordstr\"om black hole by the quantum   tunneling of charged particles. We consider the   scattering of spin-0 and
 spin-$\frac{1}{2}$ particles by the black hole in a unified framework and obtain analytically, for the first time, the pertinent reflection and transmission coefficients without any small charge approximation. Based on these results, we propose some gedanken experiments that could lead to the violation of the weak cosmic censorship conjecture due to the (classically forbidden) absorption of small energy charged particles by the black hole. As for the case of  scattering in Kerr spacetimes,
 our results demonstrate explicitly
 that scalar fields are subject to
 (electrical) superradiance phenomenon, while spin-$\frac{1}{2}$ fields are not.
 Superradiance impose some limitations on the gedanken experiments involving
 spin-0 fields, favoring, in this way,  the mechanisms for
 creation of a naked singularity by the quantum tunneling of spin-$\frac{1}{2}$ charged fermions.
 We also discuss the implications that vacuum polarization effects and quantum
 statistics might have on these gedanken experiments. In particular, we show that they are not enough to prevent the absorption of incident small energy particles and, consequently, the formation
 of a naked singularity.
\end{abstract}

\pacs{04.70.Dy, 04.20.Dw, 04.62.+v}

\maketitle

\section{Introduction}

The weak cosmic censorship conjecture (WCCC),
proposed by Roger Penrose in 1969, asserts that
any singularity originated from gravitational collapse must be hidden inside the event horizon of a black hole \cite{penrose}. Even though the WCCC is a necessary condition to guarantee the predictability of classical General Relativity, its validity remains an open question.
Since it has been proposed, numerous classical results have given strong support to the conjecture \cite{waldgedanken, boulware, *needham, *wald}. For example, if the WCCC is false, then it would be natural to expect that the formation of a black hole in gravitational collapse would be a non-generic outcome. However, the fact that stationary black holes are stable under linear perturbations provides good evidence in favor of the conjecture.  Another typical test of the WCCC consists on gedanken experiments trying to destroy the event horizon of a black hole and consequently exposing its inner singularity to an outside observer. These thought experiments are based on the uniqueness theorems that assert that all stationary black hole solutions of the Einstein-Maxwell equations are characterized by three conserved parameters: the gravitational mass $M$, the electric charge $Q$ and the angular momentum $J$, which satisfy (we assume geometrized units, so that $c=G=\hbar = 1$)
\beq \label{cond1}
M^2 \geq  Q^2 + \left( \frac{J}{M}  \right)^2.
\eeq
Solutions whose parameters do not satisfy (\ref{cond1}) are generically
dubbed naked singularities.
 When the equality holds, the black hole is called extreme. We call a black hole near-extreme if $0<M^2 -  Q^2 - \left( {J}/{M}  \right)^2 \ll M^2$.

The basic idea behind the most common  {gedanken} experiments is to make a black hole absorb a test particle with enough charge and/or angular momentum so that condition (\ref{cond1}) ceases to be valid. Wald  \cite{waldgedanken} has shown that, if the black hole is initially extreme, its event horizon is always preserved because the particle total energy required to surpass the potential barrier (created by the interaction particle-hole) more than compensates the increase in charge and/or angular momentum. We can also conclude from Wald's pioneering  work that quasi-stationary processes cannot be used to destroy the event horizon of a black hole since, in such processes, the black hole would be arbitrarily close to extremality before becoming a naked singularity. Some other classical analysis, however, indicate that a violation of the WCCC is possible if we start with a near-extreme black hole but use   non-stationary processes instead \cite{hubeny, hodselfenergy, jacobson,santa} (see also Ref.~\cite{extra1,*extra2,*extra6,*extra5,*extra3,*extra8}). Since the destruction of the event horizon, in these cases, occurs due to feeble effects beyond the linear first approximation, a way to avoid the formation of naked singularities might be to consider higher order effects, {\em e.g.} self-energy corrections of the particle's total energy \cite{hubeny, jacobson, hodselfenergy, hod1, extra4, *extra7}.

Some interesting recent attempts to destroy the event horizon are based on quantum tunneling of particles into near-extreme black holes. The original idea, by Matsas and Silva \cite{matsas}, is based on the fact that, due to the wave-particle duality, the tunneling probability can be non-zero even if the particle total energy is smaller than the height of the potential barrier, indicating a possible violation of the WCCC. However, since we lack a complete description of the quantum gravity regime, a final conclusion cannot yet be obtained, see Ref.~\cite{matsas, extra7, hod1, hod2, mauricio1, mauricio2} for
further details and considerations.

So far, only neutral particles have been treated from this quantum perspective. In this paper, we will consider the tunneling process of charged spin-0 and
 spin-$\frac{1}{2}$ particles into a charged black hole. We set up
 a unified framework for describing scattering processes of these particles
 by Reissner-Nordstr\"om ($J=0$) black holes and
  evaluate analytically the pertinent transmission and reflection coefficients for all possible values of charge. To the best of our knowledge, this has only been done before in the limit of small charges, see Ref.~\cite{gibbons, sampaio}. The calculations are performed in the
  limit of interest for violating the WCCC: the limit of small energies for
  incident massless particles.
  Based on these scattering processes, we propose some gedanken experiments that would be able, in principle, to violate the WCCC due to
 quantum tunneling of charged particles with very small energies (not sufficient to preserve (\ref{cond1}) after the absorption by the black hole).
  Interestingly enough, we have, as for the case of Kerr spacetimes, that
  spin-$\frac{1}{2}$ fermions are not
  subject to the (electrical) superradiance phenomenon, suggesting  that such
  fermions  are the best candidates for the creation of naked singularities
 in gedanken experiments involving
  quantum
  tunneling processes. In fact, we will show that superradiance   effectively
  makes more difficult the formation of a naked singularity by the absorption of a scalar
  charged particle by the black hole.
    We also consider some
  vacuum polarization and quantum statistics effects that might influence these processes, and we show that they are not enough to prevent the formation
 of a naked singularity.

\section{Charged fields around a Reissner-Nordstr\"om black hole}

A non-rotating charged black hole of mass $M$ and charge $Q$  corresponds to the Reissner-Nordstr\"om (RN) spacetime. By using spherical coordinates, the RN metric can be cast as
\beq
\label{RN}
ds^2 = - \frac{\Delta}{r^2} dt^2 + \frac{r^2}{\Delta} dr^2 + r^2 \left( d \theta ^2 + \sin ^2 \theta  d \varphi ^2 \right),
\eeq
where
\beq
\Delta = r^2 - 2Mr + Q^2 =  (r - r_+ )(r-r_- ).
\eeq
The roots of $\Delta$, denoted by $r_+$ and $r_-$,  are, respectively, the event horizon and the Cauchy horizon of the black hole. Notice that the
extreme case corresponds to $r_+=r_-$, whereas for a naked singularity,
{\em i.e.} solutions for which condition (\ref{cond1}) does not hold,
$\Delta$ has no real root.
The only non-vanishing
component of the electromagnetic potential $A_ \mu$ for the RN metric is
$A_0= -Q/r$.
Charged spin-$0$ and spin-$\frac{1}{2}$ fields
on the RN metric (\ref{RN}) can be described effectively in the unified framework that we present here.

\subsection{Massless charged scalar field}

A massless charged scalar field propagating in the RN background is described by the usual   Klein--Gordon equation \cite{nakamura},
\beq
\left(\nabla _ \mu - i q A_ \mu \right) \left(\nabla ^ \mu - i q A ^ \mu \right) \phi  = 0,
\eeq
where  $q$ stands for the electric charge of the scalar field. Note that the usual covariant derivative operator has been replaced by $\nabla_ \mu - i q A_ \mu$ to include the minimal coupling between the RN electromagnetic potential and the charge of the field. This equation can be easily separated in the RN metric if we consider the ansatz
\beq
\phi = e^{- i \omega t} e ^{i m \varphi} S_0 (\theta) R_0 (r).
\eeq
The resulting radial and angular equations will be given by the forthcoming master equations (\ref{master}) and (\ref{masterangular}) with $s=0$.

\subsection{Massless charged spin-$\frac{1}{2}$ field}

In the Newman--Penrose formalism, the wave function describing  a massless charged spin-$\frac{1}{2}$  field is represented by a pair of spinors, $P^A$ and $\bar{Q}^A$, which obey the Dirac equations
\begin{align}
\sigma^ \mu {}_{AB'} \left( \nabla _\mu - i q A_\mu \right) P^A  &= 0, \label{dirrn1} \\
\sigma^ \mu {}_{AB'} \left( \nabla _\mu + i q A_\mu \right) Q^A  &= 0, \label{dirrn2}
\end{align}
where $\sigma^\mu {}_{AB'}$ stands for the generalizations of the Pauli spin matrices called Infeld--van der Waerden symbols \cite{stewart}. Minimal coupling is also assumed in this case.   The following transformations,
\begin{align}
P^0 &= \frac{ R_{-\frac{1}{2}}(r)S_{-\frac{1}{2}}(\theta)}{ r } e^{- i \omega t} e ^{i m \varphi} , \\
P^1 &= R_{+\frac{1}{2}}(r) S_{+\frac{1}{2}}(\theta)e^{- i \omega t} e ^{i m \varphi} , \\ 
\bar Q ^{1'} &= R_{+\frac{1}{2}}(r) S_{-\frac{1}{2}}(\theta)e^{- i \omega t} e ^{i m \varphi} , \\
\bar Q ^{0'} &= - \frac{ R_{-\frac{1}{2}}(r)S_{+\frac{1}{2}}(\theta)}{r} e^{- i \omega t} e ^{i m \varphi} ,
\end{align}
can be used to separate the Dirac equations (\ref{dirrn1}) and (\ref{dirrn2}). The resulting expressions are \cite{lee, pagex, *pagey}
\begin{align}
\Delta ^{\frac{1}{2}} \left( \partial _r - i \frac{K}{\Delta} \right) R_{-\frac{1}{2}} &= \lambdabar  \sqrt{\frac{\Delta}{2}} R_{+\frac{1}{2}}, \label{119rn} \\
\Delta ^{\frac{1}{2}} \left( \partial _r + i \frac{K}{\Delta} \right) \left( \sqrt{\frac{\Delta}{2}} R_{+\frac{1}{2}} \right) &= \lambdabar R_{-\frac{1}{2}}, \label{120rn}
\end{align}
for the radial part
and
\begin{align}
\left( \partial _{\theta} + m \csc \theta + \frac{1}{2} \cot \theta \right) S_{+\frac{1}{2}} &= - \lambdabar  S_{-\frac{1}{2}}, \label{121rn} \\
\left( \partial _{\theta} - m \csc \theta + \frac{1}{2} \cot \theta \right) S_{-\frac{1}{2}} &= \lambdabar  S_{+\frac{1}{2}}, \label{122rn}
\end{align}
for the angular one,
where   $\lambdabar$ is a separation constant and
\begin{equation}
\label{KK}
K = \omega r ^2 - q Q r.
\end{equation}
It is possible to eliminate $R_{+\frac{1}{2}}$ (or, equivalently,  $R_{-\frac{1}{2}}$) from the equations ($\ref{119rn}$) and ($\ref{120rn}$), in order to obtain an equation for $R_{-\frac{1}{2}}$ (or $R_{+\frac{1}{2}}$) only. A similar procedure can be made for the angular functions. The resulting equations are given by the master equations (\ref{master}) and (\ref{masterangular}) below,  with $s=\pm 1/2$.

\section{The Scattering}

The previous radial and angular equations for spin-$0$ and spin-$\frac{1}{2}$ particles can be written in a unified manner as single master equations (analogous to the Teukolsky equations for uncharged massless fields in the Kerr metric \cite{teukolsky}). For the radial part, one has
\begin{widetext}
\begin{align}
  \Delta^{-s} \frac{d}{dr} \left(\Delta^{s+1}  \frac{dR_{s} }{dr}  \right)  +
\left(
\frac{K^2 -  2is (r-M)K}{\Delta}   + 4is \omega r - 2isqQ - \lambda _s \right)R_s   =0, \label{master}
\end{align}
where $\lambda_s$ is a separation constant, $\omega$ is the energy of the particle and $K$ is given by (\ref{KK}), while for the angular part we have
\begin{align}
\frac{1}{\sin \theta} \frac{d}{d \theta} \left( \sin \theta \frac{dS_{s}}{d \theta} \right) + \left(   s + \lambda _s- \frac{m^2}{\sin ^2 \theta}   - \frac{2 m s \cos \theta}{\sin ^2 \theta} - s^2 \cot ^2 \theta  \right) S_{s} = 0.  \label{masterangular}
\end{align}
Analogously to the Teukolsky equations, the fact that the solutions of the angular equation (\ref{masterangular}) must be regular at $\theta = 0$ and $\theta = \pi$, transforms (\ref{masterangular}) into a Sturm-Liouville problem for the separation constant $\lambda_s$ (which differs from the separation constant $\lambdabar$). The eigenfunctions which solve the problem are the spin-weighted spherical harmonics ${}_sY^m_{j}$, with corresponding eigenvalues given by $\lambda _s = (j - s)(j + s + 1)$ \cite{goldberg}. The quantities $-\left|\ell - |s|\right| \le j \le \ell + |s|$, $\ell$, and $- \ell \le m \le \ell$ are, respectively, the total, orbital and azimuthal angular momenta of the particle. Defining a new (tortoise) radial coordinate $r_*$ by $dr_* / dr = r^2 / \Delta$ and a new function $Y_s=\Delta ^{s/2}rR_{s}$, the radial equation (\ref{master}) can be written as
\beq \label{eq3}
\frac{d^2Y_s}{dr_*} + V_s(r_*)Y_s=0,
\eeq
where
\begin{align}
V_s(r_*)=\frac{\Delta}{r^4}\left[ \frac{\left( K - i s (r - M) \right) ^2}{\Delta} + 4 is \omega r   - 2 isqQ - j(j+1) + s^2 - 2 \frac{M}{r} + 2 \frac{Q^2}{r^2}  \right].
\end{align}
\end{widetext}
Let us now consider   the scattering of an incident wave originated far away from the black hole. This process can be described by a solution $Y_s$ of the wave equation with the following asymptotic behavior far from the black hole (see the Appendix A
for the calculation details)
\beq \label{asymp1}
Y_s \rightarrow Z_s^{\rm in} r_* ^{s + i q Q} e^ {- i \omega r_*} + Z_s^{\rm out} r_* ^{-s - i q Q} e^ {+ i \omega r_*},
\eeq
and the following asymptotic behaviour near the event horizon of the black hole,
\beq \label{asymp2}
Y_s \rightarrow Z_s^{\rm tr} e^{-\frac{s}{2}\left(\frac{r_+ - r_-}{r_+ ^2} \right)r_* - i \left( \omega - \frac{q Q}{r_+} \right) r_ * }.
\eeq
The $Z_s^{\rm in}$, $Z_s^{\rm out}$ and $Z_s^{\rm tr}$ coefficients
 correspond, respectively,
 to the incident, reflected and transmitted wave amplitudes.
   Note that, in the fermionic case, $Z_{+\frac{1}{2}}$ and $Z_{-\frac{1}{2}}$ are not independent. For such a case, plugging the asymptotic forms (\ref{asymp1}) and (\ref{asymp2})  into Eqs.~(\ref{119rn}) and (\ref{120rn}), we obtain
\begin{align}
2i \omega Z_{+\frac{1}{2}}^{\rm out} &= \lambdabar \sqrt{2} Z_{- \frac{1}{2}} ^{\rm out}, \label{xx1}\\
2 \sqrt{2} i \omega Z_{-\frac{1}{2}} ^{\rm in} &= -\lambdabar Z_{+ \frac{1}{2}} ^{\rm in}, \label{xx2}\\
\frac{\lambdabar}{\sqrt{2}} Z_{+\frac{1}{2}} ^{\rm tr}  &= \left(\frac{1}{2} - 2 i \delta \right)\left( r_+ - r_- \right) ^{1/2} Z_{-\frac{1}{2}} ^{\rm tr}, \label{xx3}
\end{align}
where $\delta$ is given by
\beq
\delta = \frac{r_+ ^2}{r_+-r_-}\left(\omega - \frac{qQ}{r_+} \right).
\eeq
In order to derive relations between the (in/out/tr) coefficients, we use the fact that the Wronskian $W$ of two linearly independent solutions of the radial equation
is independent of $r$,
\beq \label{wronsk}
  W[Y_s,Y^*_{-s}] _{r=r_+} =   W[Y_s,Y^*_{-s}] _{r=\infty}.
\eeq
Using the solution whose asymptotic behaviour is given by Eqs.~(\ref{asymp1}) and (\ref{asymp2}), we obtain the relation
\beq
r_{s}+t_{s}=1,
\eeq
where
\beq \label{spin0}
r_{0} = \left| \frac{Z_0 ^{\rm out}}{Z_0 ^{\rm in}} \right| ^2, \quad t _{0} = \frac{1}{\omega} \left(\omega - \frac{q Q}{r_+} \right)\left| \frac{Z_0 ^{\rm tr}}{Z_0 ^{\rm in}} \right| ^2,
\eeq
for the spin-$0$ case and
\beq \label{spin12}
r_{\frac{1}{2}} = 4 \frac{\omega ^2}{\lambdabar ^2} \left| \frac{Z_{+ \frac{1}{2}} ^{\rm out}}{Z_{+ \frac{1}{2}} ^{\rm in}} \right| ^2, \quad t_{\frac{1}{2}} = \frac{\sqrt{r_+ - r_-}}{r_+ ^2}  \left| \frac{Z_{+ \frac{1}{2}} ^{\rm tr}}{Z_{+ \frac{1}{2}} ^{\rm in}} \right| ^2,
\eeq
for the spin-$\frac{1}{2}$ case
(similar expressions for the case $s=-\frac{1}{2}$ can be obtained by substituting equations (\ref{xx1})-(\ref{xx3}) into the expressions above).
The quantities $r_{s}$ and $t_{s}$ can be interpreted as reflection and transmission coefficients, as confirmed by calculations of energy fluxes \cite{lee}. (See the Appendix A for the pertinent calculations.) We can see that $t_{0}$ can assume negative values when $\delta < 0$, {\em i.e.},
when
\beq
\omega - \frac{qQ}{r_+} <0.
\eeq
Therefore, similarly to what happens for an uncharged scalar particle scattered by a Kerr black hole \cite{super1, *super2}, a charged particle scattered by  a RN black hole can effectively extract energy from it \cite{bekenstein}. This phenomenon, generically called superradiance \cite{super3,*mauricio3}, can be interpreted as a stimulated emission process \cite{stimulated} for certain particles. Note that the transmission coefficient for  spin-$\frac{1}{2}$ fermions is always non-negative and, therefore, (electrical) superradiance is also impossible for such fermions \cite{lee}, exactly in the same manner
it occurs for the Kerr case  \cite{super4}.

\section{Challenging the WCCC}

We are now ready to investigate the validity of the WCCC when low energy spin-$0$ and spin-$\frac{1}{2}$  particles tunnel into a near-extreme RN black hole. Before the scattering process starts, the black hole's mass $M$ and charge $Q$ satisfy the relation
\beq
0 < M^2 - Q^2 \ll M^2.
\eeq
As a result of the tunneling process, the black hole absorbs the particle total energy $\omega$, charge $q$ and the total angular momentum $L$. If we want to destroy the event horizon and create a naked singularity, the following relation must be satisfied
\beq \label{wccccond}
f=(M+\omega)^2 - (Q+q)^2 - \frac{L^2}{(M+\omega)^2} < 0.
\eeq
In order to make this process possible even for zero angular momentum particles, we choose $Q$ and $q$ to have the same sign, which we assume, without loss of generality for the purposes here, to be positive. (The   calculations for the general
 case are presented in the Appendix.) In the spinless case, the particle's energy cannot be arbitrarily small, otherwise superradiance occurs and, instead of tunneling into the black hole, the particle extracts energy and charge from it \cite{bekenstein}. From Eq.~(\ref{spin0}), we conclude that $\omega$ must satisfy
\beq \label{cond_rncar1}
\omega > \frac{q Q}{ r_+}
\eeq
if the particle is to be absorbed by the black hole. Choosing the incident particle to have zero angular momentum ({\em i.e.} $\ell = j = 0$ and, therefore, $L=0$) in order to minimize any backreaction effect related to rotations, condition (\ref{wccccond}) above reduces simply to
\beq \label{cond_rncar2}
\omega < q - (M- Q).
\eeq
Note that, similarly to what happens in the classical process \cite{hubeny}, if the black hole is initially extreme, {\em i.e.} $M=Q$, we have $r_+=M$ and
it is impossible to find $\omega$ satisfying Eqs.~(\ref{cond_rncar1}) and (\ref{cond_rncar2}) simultaneously. However, for a near-extreme black hole, if the particle's charge is chosen to be
\beq
\label{min}
q > \frac{M - Q}{r_+ - Q} r_+ = \frac{r_+-Q}{2},
\eeq
then it is always possible to find energies $\omega$ for which the inequalities (\ref{cond_rncar1}) and (\ref{cond_rncar2}) are satisfied simultaneously.
Notice that the restriction (\ref{min}) is exactly the same obtained previously for the classical
gedanken experiments involving charged test bodies, see \cite{hubeny, santa}. For the limit
of near extreme black holes, one has
\beq
\label{minn2}
q  > q_0 = \sqrt{\frac{M \varepsilon}{2}} + O(\varepsilon),
\eeq
where $M - Q = \varepsilon > 0$.
Let us consider now $\omega =  ({qQ}/{r_+}) + \delta_0$, with positive $\delta_0$. Expanding the expression for $f$ in the lowest orders for $\delta_0$ and $\varepsilon$, we obtain
\beq \label{eqq2}
f = 2 (M+q) \delta_0 - 2 q\sqrt{\frac{2 {\varepsilon}}{M}}     (M+q)   +  O(\delta_0 ^2,\varepsilon).
\eeq
We conclude that if $\delta_0$ and $\varepsilon$ are sufficiently small and satisfy
\beq \label{cond_delt}
0 < \delta_0 < q\sqrt{\frac{2 {\varepsilon}}{M}} ,
\eeq
with $q>q_0$,
then the condition (\ref{wccccond}) for the creation of a naked singularity is fulfilled.
It is important we keep in these absorption processes the validity of the
test field approximation, $\omega \ll M$ and $q \ll Q$, otherwise a
myriad of backreaction effects could appear   preventing effectively
the absorption. Even though it is possible to keep the deviations of the
test field approximation for scalar fields to a minimum, we will see that
the situation for the  fermions is still more favorable.

For the spin-$\frac{1}{2}$ case, superradiance is absent, and the particle's energy $\omega$ can, in principle, have any value greater than zero that the tunneling probability will remain positive even for arbitrarily small $q$, keeping
in this way backreaction effects really to the minimum.
  Let us consider again  $\ell = 0$ (and, therefore, $L^2 = s(s+1)=3/4$). The first order expansion of $f$ in terms of $\omega$ and $\varepsilon$ is given by
\beq
f \approx 2\left( M + \frac{3}{4 M^3} \right) \omega + 2 (M+ q ) \varepsilon - 2 M  q  - q^2 - \frac{3}{4 M^2}.
\eeq
The condition necessary  to impose Eq.~(\ref{wccccond}) and, consequently, create a naked singularity is
\beq
\label{naked}
0< \omega < \frac{2 M  q  + q^2 + \frac{3}{4 M^2} - 2 (M+ q ) \varepsilon}{2M + \frac{3}{2 M^3}}.
\eeq
In contrast with the scalar case, there is no minimal value for $q$ in order to assure the tunneling. Let us suppose that the spin-$\frac{1}{2}$ field be a
small perturbation (both $\omega,q \ll M$) and that the black hole is large ($M\gg 1$, assuring in this way that total angular momentum $L^2 = 3/4$ of the fermion is also a small
perturbation). In this approximation, Eq.~(\ref{naked}) reads
\beq
\label{naked1}
0 < \omega < q - \varepsilon,
\eeq
whose interpretation is rather simple: the only requirement on the charge $q$ is that it must be enough to overcharge the black hole. For any charge capable of this, there are energies for which the incident particles might convert the black hole into a naked singularity. Spin-$\frac{1}{2}$ fields can   elude the backreaction effects arising from the breakdown of the test field approximation in the most
efficient way.

\subsection{Vacuum polarization and quantum statistics}

Notwithstanding,
there is a
  point of concern for fermions, raised by Hod in Ref.~\cite{hod2}: the spontaneous polarization of the vacuum around a black hole and the Pauli exclusion
  principle. In order to properly introduce such  effect, let's briefly review some aspects of  particle production by black holes. In time-dependent spacetimes, {\em e.g.} the gravitational collapse of a spherical body, particle production occurs because the Hamiltonian describing the evolution of the field is time-dependent and, therefore a mixing of positive and negative frequencies can happen. Since particles and antiparticles are usually described by positive and negative frequencies, one then says that particle production has occured. However, the correct definition of particle/antiparticle states is given in terms of an appropriate pseudonorm (in order to be compatible with the (anti-)commutation relations). For example, for scalar fields the pseudonorm is naturally defined from the Klein--Gordon scalar product. When a mode has a positive norm, it is called a particle state; on the other hand, when a mode has negative norm, it is called an antiparticle state. Usually, this definition is equivalent to defining particle states with respect to the corresponding mode frequency. However, for charged black holes (and also for rotating ones) this is not the case. The eletromagnetic interaction between the field and the black hole makes it possible for a mode with given frequency $\omega$ to have positive norm in the vicinity of the black hole and negative norm far away from it, characterizing again particle production. In other words, the electric field is so strong in the vicinity of the black hole that a pair particle/antiparticle can be created \cite{gibbons, kim}. One of them escapes to infinity, since it has positive energy (as measured by an observer at infinity) and its charge has the same sign as the black hole's charge. The other particle (antiparticle), has negative energy and is absorbed by the black hole, discharging it and reducing its electromagnetic energy. The expected number of particles spontaneously emitted in each mode is \cite{gibbons,bekenstein_ab}
\begin{equation}
\langle n_{\frac{1}{2}}\rangle = t_{\frac{1}{2}} \left[ 1 + e^x \right]^{-1},  \ \ \ x=\frac{\omega - q Q / r_+}{T_{\rm bh}},
\end{equation}
where $T_{\rm bh}$ is the temperature of the black hole and the transmission coefficient $t_{\frac{1}{2}}$ is given by expression (\ref{tra_coe}). For extreme black holes ($T_{\rm bh} = 0$) only superradiant modes $\left(\omega < q Q / r_+ \right)$ are emitted. Back to the question of cosmic censorship, the interesting situation to be considered is that of a black hole immersed in a thermal radiation bath which obeys Fermi--Dirac statistics. Including spontaneous emission and pure scattering effects, the probability that one fermion is incident on the black hole without reflection is \cite{bekenstein_ab,wald_prob}
\beq \label{prob12}
p_{\frac{1}{2}} = t_{\frac{1}{2}} \left[ 1 + e^{-x} \right]^{-1}.
\eeq
 Clearly, if the black hole is initially extreme, the probability that a low energy particle ($\omega < q Q /r_+ $) enters the black hole vanishes. However, for a near-extreme black hole, the probability is, albeit small, always positive and, therefore, the possibility of creating a naked singularity really exists.
Consider, for example, the case of a very low energy ($\omega M \ll 1$) and low angular momentum ($\ell \ll qM$) particle, incident on a near extreme black hole. Eq.~(\ref{prob12}) reduces, in this limit, to
\beq
p_{\frac{1}{2}} \rightarrow e^{x},
\eeq
where expression (\ref{extreme1}) for the corresponding transmission coefficient $t_{\frac{1}{2}}$ was used.
   Note the similarity of this result with the classical gedanken experiments described in the introduction, in which a violation of the WCCC can only happen for near-extreme black holes (and not for extreme black holes).

	It is also interesting  to analyze the effect of vacuum polarization on the tunneling probability of scalar particles by considering a charged black hole immersed in a thermal radiation bath obeying Bose--Einstein statistics. In this case, including spontaneous emission, stimulated emission and pure scattering effects, the probability that $n$ scalar particles are incident on the black hole without reflection is \cite{bekenstein_ab,wald_prob}
\begin{equation} \label{prob0}
p_0 = t_0 ^n e^{xn} \frac{e^x - 1}{\left(e^x - 1 + t_0\right)^{n+1}},
\end{equation}
where $t_0$ is given by expression (\ref{tra_coe}). Since $p_0$ is always positive for non-extreme black holes, even superradiant modes have a non-zero probability of being absorbed without any kind of emission/reflection by the black hole. Therefore, conditions (\ref{cond_rncar1}) and (\ref{cond_rncar2}) are  replaced by
\begin{equation}
0 < \omega < nq - (M - Q),
\end{equation}
and we conclude that it is possible, in principle, to violate the WCCC using
scalar particles with small charges $q$, even though the typical probabilities
 can be considerably small.
 In particular, for a very low energy ($\omega M \ll 1$) and low angular momentum ($\ell \ll qM$) particle, Eq.~(\ref{prob0}) reduces to
\beq
p_{0} \rightarrow 2^{-(n+1)}e^{xn},
\eeq
where expression (\ref{extreme1}) for the corresponding transmission coefficient $t_{0}$ was used.
Observe that one could, effectively,  use  several particles ($n>1$) with $q<\varepsilon $ in order to create a naked singularity. However, the greatest probability (same order of magnitude of the spin-$\frac{1}{2}$ case) is obtained for a single particle ($n=1$).

\section{Conclusion}

We investigated the possibility of violating the WCCC by quantum tunneling of charged spin-$0$ and spin-$\frac{1}{2}$ particles into a RN black hole. If the black hole configuration is initially extreme $M=Q$, the tunneling of particles with enough charge to create a naked singularity is impossible, irrespective of the
 particle spin. This was also the case in classical  {gedanken} experiments attempting to destroy the event horizon of a black hole \cite{waldgedanken, hubeny, jacobson}. However, for a near-extreme RN black hole, an incident particle can, in principle, be quantum mecanically absorbed and overcharge the black hole, destroying its event horizon. Besides, we obtained analytical results, in the small energy limit ($M \omega \ll 1$), for the transmission coefficient of spin-$0$ and spin-$\frac{1}{2}$  fields scattered by a charged black hole. As for the Kerr spacetime case, electrical superradiance makes more difficult the formation of a naked
 singularity by the absorption of scalar particles. The situation for the scalar field is similar to the classical gedanken experiments involving small charged bodies \cite{hubeny, santa}.
 However, for  spin-$\frac{1}{2}$, superradiance is absent, and it is possible to choose the parameters of the particle in order to destroy the black hole's event horizon keeping to a minimum any backreaction effect related to the breakdown of the test field
 approximation. We also showed that vacuum polarization effects cannot be invoked to
 elude the creation of a naked singularity.	

The quantum nature of the processes we are proposing here is fundamental for the violation of the WCCC. Indeed, Ref.~\cite{mauricio2} shows, in a very similar context, how different the quantum tunneling and its classical limit can be. Even though the particle propagates as a wave due to the wave-particle duality, it tunnels into the black hole as a single quantum.
For instance,
a possible way to include backreaction effects for scalar particles would be to solve the coupled Einstein--Klein--Gordon equations,
\begin{align}
\left(\nabla _ \mu - i q A_ \mu \right) \left(\nabla ^ \mu - i q A ^ \mu \right) \phi  = 0 &, \\
G_{\mu \nu} = 8 \pi \langle T_{\mu \nu} \rangle &,
\end{align}
 where $\langle T_{\mu \nu} \rangle$   is  the vacuum expectation value for the stress energy tensor naturally assigned to the quantum scalar  field $\phi$. However, the change from a black hole configuration to a naked singularity is certainly not a smooth process. Even this backreaction calculation would be insufficient to determine the fate of the black hole. A definite answer about the validity of the WCCC will only be possible when a complete quantum gravity theory becomes available.

\acknowledgments
This work was supported by FAPESP and CNPq. The authors are grateful to G.
Matsas, R.P. Macedo, R. Santarelli, T.P. Sotiriou, and D. Vanzella for enlightening discussions.

\appendix

\section{Reflection and transmission coefficients}

In order to determine the
reflection and transmission coefficients (\ref{spin0}) and
 (\ref{spin12}) in the small energy limit $M\omega \ll 1$ of the radial
 master equation (\ref{master}), we follow the same approach used in
 \cite{starobinski, *page3} for the Kerr case. The charges $q$ and $Q$ here
 are generic. By introducing
\beq
x = \frac{r-r_+}{r_+-r_-},
\eeq
Eq.~(\ref{master}) can be written in the limit $M \omega \ll 1$ as
\begin{widetext}
\beq \label{master2}
x^2 (x+1)^2 \frac{d^2 R_s}{dx^2} + (s+1)(2x + 1)x(x+1) \frac{dR_s}{dx} + \left[ k^2 x^4 + 2\left(is -qQ \right) k x^3 + (q^2Q^2 - \lambda _s)x(x+1) + G_s x + H_s \right] R_s = 0,
\eeq
\end{widetext}
where
\begin{align}
k  &= \omega \left( r_+ -r_- \right), \\
G_s &= \left(\frac{s}{2} - i \delta - i q Q \right)^2 - \left(\frac{s}{2} + i \delta \right)^2, \\
H_s &= \left(\frac{s}{2}  \right)^2 - \left(\frac{s}{2} + i \delta \right)^2.
\end{align}
In the limit $kx \ll 1$, the first two terms inside the square brackets in Eq.~(\ref{master2}) can be ignored. The solution corresponding to an ingoing wave near the horizon is given by
\beq \label{solution1}
R_s= A_s (1+x)^{-s + i \delta + i q Q} x^{-s - i \delta} \, {} _2 F _1 \left(a,b;c;-x\right),
\eeq
where $_2 F _1(a,b;c;x)$ stands for the ordinary hypergeometric function,
\beq
a = \frac{1}{2} - s + iqQ - i \beta, \quad b = \frac{1}{2} - s + iqQ + i \beta,
\eeq
\beq
c = 1 - s - 2i \delta , \quad \beta = \sqrt{q^2 Q^2 - \left(j + \frac{1}{2}\right)^2},
\eeq
and $A_s$ must be chosen so that Eqs.~(\ref{xx1})-(\ref{xx3}) are satisfied. In fact,
comparing the asymptotic behavior near the horizon of Eq.~(\ref{solution1}) with Eq.~(\ref{asymp2}), we have
\beq \label{zstr}
Z_s^{\rm tr} =A_s \, r_+ \left(r_+ - r_-\right) ^{\frac{3}{2}s + i \delta}.
\eeq
If Eq. (\ref{xx3}) is to be satisfied, we conclude that $A_{+\frac{1}{2}}$ and $A_{-\frac{1}{2}}$ must be related according to
\beq
\frac{A_{+\frac{1}{2}}}{A_{-\frac{1}{2}}} = \frac{\sqrt 2}{\lambdabar \left(r_+ - r_- \right)} \left(\frac{1}{2} - 2 i \delta \right).
\eeq
Once one of the parameters is arbitrarily chosen, the other one is automatically determined by the expression above. For the spin-$0$ case, there is only one parameter, $A_0$, which can be chosen arbitrarily.  

Considering now the limit $x \gg 1$, the last two terms inside the square brackets in Eq.~(\ref{master2}) can be dropped and $x+1$ can be simply replaced by $x$. The corresponding solution is
\begin{widetext}
\begin{align}
 R_s = C_1 x^{-\frac{1}{2} -s + i \beta}e^{-ikx} \, {}_1F_1 \left(d_1,1 + 2i\beta; 2ikx \right)
 + C_2 x^{-\frac{1}{2} -s - i \beta}e^{-ikx} \, {}_1F_1 \left(d_2,1 - 2i\beta; 2ikx \right), \label{solution2}
\end{align}
where $_1 F _1(a,b;x)$ stands for the confluent hypergeometric function and $d_1$ and $d_2$ are given by
\beq
d_1 = \frac{1}{2}-s+i\beta -iqQ, \qquad d_2 = \frac{1}{2}-s-i\beta -iqQ.
\eeq
By matching the two solutions, {\em i.e.} Eqs.~(\ref{solution1}) and (\ref{solution2}), in the overlap region $1 \ll x \ll 1/k$,  it is possible to determine the coefficients $C_1$ and $C_2$,
\beq
C_1 = A_s \frac{\Gamma(2i \beta) \Gamma(c)}{\Gamma(b) \Gamma(c-a)}, \quad C_2 = A_s \frac{\Gamma(-2i \beta) \Gamma(c) }{\Gamma(a) \Gamma(c-b)}
\eeq
Now, comparing the asymptotic form of the confluent hypergeometric functions $_1 F _1(a,b;x)$ with Eq.~(\ref{asymp1}), we find
\begin{align}
\frac{Z_s^{\rm in}}{(r_+-r_-)^{1-iqQ}} = C_1 \frac{ \Gamma(1+2i \beta) (-2ik)^{-\frac{1}{2} + s - i \beta +iqQ}}{\Gamma \left(\frac{1}{2} + s + i\beta +i q Q \right) }  + C_2 \frac{ \Gamma(1-2i \beta) (-2ik)^{-\frac{1}{2} + s + i \beta +iqQ}}{\Gamma \left(\frac{1}{2} + s - i\beta +i q Q \right) }. \label{zsin}
\end{align}
Using Eqs.~(\ref{zstr}) and (\ref{zsin}), we calculate
\beq \label{ztrzin}
 \left| \frac{Z_{s} ^{\rm tr}}{Z_{s} ^{\rm in}} \right| ^2 = \frac{r_+ ^2 (r_+-r_-)^{3s - 2}e^{-\pi q Q} (2k)^{1-2s}}{|\Gamma(c)|^2 |F(\beta) + F(- \beta)|^2},
\eeq
where
\beq
F(\beta) = \frac{\Gamma(2i\beta) \Gamma(1 + 2i \beta) e^{-\frac{\pi \beta}{2} - i\beta \log(2k)} }{\Gamma(b)\Gamma(c-a)\Gamma\left( \frac{1}{2} + s + iqQ + i\beta \right)}.
\eeq
Plugging Eq.~(\ref{ztrzin}) into the expressions (\ref{spin0}) and (\ref{spin12}), we obtain the formula for the transmission coefficient,
\beq \label{tra_coe}
t_{s} = \frac{e^{- \pi q Q}(2\delta)^{1-2|s|}}{|\Gamma(1 - |s| - 2i \delta)|^2 |F(\beta) + F(- \beta)|^2}.
\eeq
When $j + 1/2 > |qQ|$, $\beta$ is purely imaginary, $\beta = i \gamma$, with $\gamma > 0$. In the limit of small energies, $k \rightarrow 0$, we can write
\beq
t_{s} = \frac{e^{-\pi q Q}(2\delta)^{1-2|s|} \left| \Gamma \left( \frac{1}{2} + |s| + \gamma + iqQ \right) \Gamma \left( \frac{1}{2} -|s| + \gamma + iqQ \right) \Gamma \left( \frac{1}{2} + \gamma - iqQ - 2 i \delta \right) \right| ^2}{\left| \Gamma(1-|s|-2i \delta) \right| ^2 \left[\Gamma (2 \gamma) \Gamma (1 + 2 \gamma) \right] ^2} (2k)^{2 \gamma},
\eeq
\end{widetext}
plus terms of order $O(k^{4\gamma})$ or higher. A similar result was obtained by Gibbons for massive scalar particles in the particular case $|qQ|\ll1$, see Ref.~\cite{gibbons}. Sampaio considered both cases of spin-0 and spin-$\frac{1}{2}$ particles, also in the same limit, in Ref.~\cite{sampaio}.  
However, when $j + 1/2 < |qQ|$, $\beta$ becomes a real number and, therefore, the transmission coefficient $t_{s}$ becomes a constant part plus an oscillating term in $k$. In the extreme case $j + 1/2 \ll |qQ|$, the constant part dominates. When the charge of the black hole and the charge of the particle have the same sign (electric repulsion), the transmission coefficient reduces to
\beq \label{extreme1}
t_{s} = (4|s|-1) + O \left(e^{-\pi qQ} \right),
\eeq
 On the other hand, when the charges have opposite signs, the transmission coefficient is given by
\beq \label{extreme2}
t_{s} = 1 - \frac{\pi}{|qQ|} \left(j + \frac{1}{2} \right) ^2 + O \left( \frac{1}{q^2 Q^2} \right).
\eeq
The fact that the transmission coefficient approaches unity in this case is not suprising since the interaction between the particle and the black hole is attractive. However, interesting results arise when $qQ>0$. First, for spin-$\frac{1}{2}$ particles, even though the electromagnetic force is repulsive, the transmission coefficient approaches unity. Second, for scalar particles the transmission coefficient approaches $-1$ (and the reflection coefficient approaches $2$), characterizing a situation of extreme superradiance.
These extreme effects are caused by the fact that the electromagnetic term is the dominant term in the coefficient $q^2 Q^2 - \lambda_s$ of $x(x+1)$ in Eq.~(\ref{master2}) and, no matter what the relative sign between the charges is, it always contributes with a positive sign.

\end{document}